% amstex -xL epsf
\input amstex
\input epsf
%\magnification=\magstep1
\documentstyle{amsppt}
\def\l{\lambda}
\def\e{\epsilon}
\def\a{\alpha}
\def\b{\beta}
\def\lmat{\left( \matrix}
\def\rmat{ \endmatrix \right)}
\def\ss{\vskip .1in}
\def\ms{\vskip .2in}
\def\ap{\rightarrow}
\def\api{\rightarrow \infty}
\def\apm{ \rightarrow -\infty}
\def\inxp{\int_x^{\infty}}
\def\inmx{\int_{-\infty}^x}
\def\inmp{\int_{-\infty}^{\infty}}
\def\ddx{\frac{d}{dx}}
\def\p{^{\prime}}
\def\pp{^{\prime \prime}}
\def\k{\kappa}
\def\mo{m^{\text{out}}}
\def\mi{m^{\text{in}}}
\def\o{\omega}
\def\sech{\text{sech}}
\def\Res{\text{Res}\,}
\TagsOnRight

\topmatter
\title A Riemann-Hilbert Problem for an Energy Dependent Schr\"odinger Operator \endtitle
\rightheadtext{}
\author David H. Sattinger
and Jacek Szmigielski \endauthor

\affil  University of Minnesota and the University of Saskatchewan\endaffil
\leftheadtext{}
\footnote[]{Research of the authors was
supported by National Science
Foundation grants  DMS-9501233 and Natural Sciences and Engineering Research Council of Canada}

\abstract{ We consider an inverse scattering problem for Schr\"odinger operators with energy dependent potentials. The inverse problem is formulated as a Riemann-Hilbert problem on a Riemann surface. A vanishing lemma is proved for two distinct symmetry classes. As an application we prove
 global existence theorems for the two distinct systems of 
partial differential equations $u_t+(u^2/2+w)_x=0,\ w_t\pm u_{xxx}+(uw)_x=0$ for suitably restricted, complementary classes of initial data.}
\endabstract

\endtopmatter
\document
\baselineskip 15pt

\centerline{ Dedicated to {\it Hugh Turrittin} on his 90th birthday}

\head 1. Introduction \endhead

In this paper the scattering theory of the energy dependent Schr\"odinger
operator
$$
(D^2+k^2)\psi=(ikp(x)+q(x))\psi,
\qquad
D=\frac{d}{dx},
\tag1.1
$$
is investigated and used to prove global existence theorems for the  ``isospectral flows" :
$$
u_t+(w+\frac{u^2}{2})_x=0,
\qquad
w_t+ u_{xxx}+(wu)_x=0;
\qquad
p=\frac{u}{2},
\qquad
q=\frac{u^2}{16}-\frac{w}{4}
\tag1.2
$$
and
$$
u_t+(w+\frac{u^2}{2})_x=0,
\qquad
w_t- u_{xxx}+(wu)_x=0
\qquad
p=\frac{iu}{2},
\qquad
q=\frac{w}{4}-\frac{u^2}{16}.
\tag1.3
$$

The scattering problem for such energy dependent Schr\"odinger operators
was first considered by Jaulent [8], Jaulent and Jean [9], 
Kaup [10], and more recently by the present authors [14]. 
In [8], [9] and [14] the problem was considered for  purely 
imaginary $p$ and real $q$ tending to zero at infinity. 
The inverse problem was solved by a modification of the 
Gel'fand-Levitan-Marchenko method. 

Kaup studied the scattering problem under the assumption of real $p$, and $q$ tending to non-zero limits at infinity, in connection with a long wave approximation of Boussinesq type (equations (1.2)). He obtained a 
coupled pair of equations of GLM type, but did not investigate their solvability.

The GLM equations are of Fredholm type, and so their solvability is a consequence of
uniqueness. Such uniqueness theorems are valid, (though usually not proved in the literature) in the case of the
standard Schr\"odinger equation, or the isospectral problem obtained by
Zakharov and Shabat in their study of the nonlinear Schr\"odinger equation [15]. But they
fail in general for  $2\times 2$
AKNS systems without further symmetry assumptions on the potential, 
for example,
that it is symmetric or skew symmetric, Hermitian or skew-Hermitian.

We investigate the solvability of the inverse scattering problem for
(1.1) in the two cases $p$ real or imaginary, corresponding to the two
flows (1.2) and (1.3).  Uniqueness theorems are proved in these
 two cases under complementary, restricted 
conditions on the scattering data. Global existence theorems for the
flows (1.2), (1.3) can then be proved by showing that the constraints
on the scattering data are invariant under the flows.

This, however, entails restrictions on the initial data.
Since global existence theorems can be proved for the two flows only
for restricted initial data, the physical relevance of these two
models is unclear.
Our interest in the matter stems rather from the novel features of the
associated inverse
scattering problem, and its formulation as a Riemann-Hilbert problem
on a Riemann surface.

Riemann-Hilbert problems arise in integrable systems as a more general
formulation of inverse scattering problems. In second order cases,
such as the Schr\"odinger and Zakharov-Shabat scattering problems,  
the Gel'fand-Levitan-Marchenko theory and Riemann-Hilbert method 
are equivalent, being essentially Fourier transforms of one another. 
But there is no GLM theory for  $n\times n$ first order systems  
or $n^{th}$ order ordinary differential operators ($n>2$); and the inverse scattering problem must be formulated as a Riemann-Hilbert problem [1, 2]. 
Moreover, even for second order problems 
the method of Riemann-Hilbert problems, coupled with the
method of steepest descent, can be used to obtain precise asymptotic
behavior
of the solutions [5].

One can go from (1.2) to (1.3) by the transformation
$$
u(x,t)\ap iu(x,it),
\qquad 
w(x,t)\ap -w(x,it);
$$
but this transformation is complex, so  
the two equations may be considered to be distinct real forms of one another.

Sachs [13] found a  pair of tau functions for (1.2) and used them to
construct rational solutions similar to the Calogero-Moser solutions of
the KdV equation. Matveev and Yavor [11] used Riemann surface theory
to construct finite gap solutions, and, in the limiting case, multi-soliton
solutions. 

Isospectral flows for energy dependent operators arise in a wide variety of applications. Recently, for example, Camassa and Holm [4] found a special shallow water wave equation generated by an energy dependent iso-spectral operator related to the Dym hierarchy. For a formal discussion of isospectral flows generated by operators with energy dependent potentials, see Fordy [7].

The  Gel'fand-Dikki hierarchy of isospectral flows of $n^{th}$ order
ordinary differential operators can be formally extended 
to operators $L$ in which the coefficient of $D^j$ is a polynomial of degree $j-1$ in the spectral parameter. 
These considerations suggest, at least on a formal level, the existence of a large class of flows, of which the Gel'fand-Dikki flows are a special case.
We conjecture that all these flows are local. The corresponding
inverse problems lead to a broader class of Riemann-Hilbert problems
than those considered in [3].

\head 2. The forward problem \endhead

Throughout this paper we assume that $q\ap 1$ as $x\ap \pm \infty$ and
rewrite (1.1) as
$$
(D^2+E^2-(ikp+q))\psi=0,
\qquad
E^2=k^2+1, \tag2.1
$$
where now $p$ and $q$ belong to Schwartz class $\Cal S$. Although
the restriction to potentials in Schwartz class could be weakened
considerably
in the theory of the forward and inverse scattering problem, the class
is a convenient one for discussing completely integrable systems with
an infinite
number of conservation laws, and so we shall make this assumption throughout.

The scattering problem may be formulated as a
Riemann-Hilbert problem on the  Riemann surface $E^2=k^2+1.$ 
Following Kaup, we introduce the uniformizing parameter $z$:
$$
E=\frac{1}{2}\left(z+\frac{1}{z}\right),
\qquad
k=\frac{1}{2}\left(z-\frac{1}{z}\right).  \tag2.2
$$
The transformation $z\mapsto E$ is a double covering of the
$E$ plane by the $z$ plane. Define four regions
$\Cal U_j, \ j=1,2,3,4,$ where $\Cal U_1=\{z:\Im\,z>0 \cap |z|>1\},$ 
$\Cal U_2$ is its image  under complex conjugation,
$\Cal U_3$ the upper half of  the unit disk, and
$\Cal U_4$ the lower half of the unit disk.
Let
$$
\Cal U_+=\Cal U_1 \cup \Cal U_4=\{z:\Im E >0\},
\qquad
 \Cal U_-=\Cal U_2 \cup \Cal U_3=\{z:\Im E <0\}.
$$
Finally, let $\Sigma$ denote the union of the real line and the unit
circle in the $z$ plane, less the origin.

We construct two sets of wavefunctions of (2.1), $\psi$ and $\phi$ defined
by the following asymptotic behavior:
$$
\psi_{\pm}(x,z) =\mo_{\pm} e^{\pm iEx},\quad z\in \Cal U_{\pm}
\qquad
\phi_{\pm}(x,z)=\mi_{\pm} e^{\mp iEx},\quad z\in \Cal U_{\pm}
$$
where $\mo_\pm$ and $\mi_\pm$ are analytic on $\Cal U_\pm$ 
respectively, and
$$
\lim_{x\apm}\mi_\pm=\lim_{x\api}\mo_\pm = 1.
$$

Equation (2.1) can be converted to a Volterra integral equation, for example,
$$
\psi_{\pm}(x,z)=e^{\pm i Ex}-\inxp \frac{\sin\, E(x-y)}{ E}(ikp(y)+q(y))\psi_{\pm}(y,z) \, dy,
\quad z\in \Cal U_+. 
$$
This leads to the following integral equations for $\mo_{\pm}$:
$$
\mo_{\pm}(x,z)=1 \mp \inxp \frac{1 - e^{\mp 2i E(x-y)}}{2i E}(ikp(y)+q(y))\mo_{\pm}(y,z)\,dy,
\qquad
z\in \Cal U_{\pm}. \tag2.3a
$$
The wave functions $\phi_{\pm}$ and $\mi_{\pm}$ satisfy similar Volterra integral equations.
For example, 
$$
\mi_{\pm}(x,z)=1 \mp \inmx \frac{1-e^{\pm 2i E(x-y)}}{2i E}(ikp(y)+q(y))\mi_{\pm}(y,z)\,dy,
\qquad
z\in \Cal U_{\pm}. \tag2.3b
$$  

The integral equations for $\mi_{\pm},\ \mo_{\pm}$ may be solved by successive approximations for all $z \in 
\overline{\Cal U_{\pm}}\setminus\{0,\,\infty\}$. (See also the form
of the wave functions given in the Appendix.)

Throughout this paper we shall understand by $\psi$ the sectionally analytic
function
$$
\psi(x,z)=\cases \psi_+(x,z) & z\in \Cal U_+; \\
\psi_-(x,z) & z\in \Cal U_-.
\endcases
$$
The other wave functions are denoted similarly. 

The asymptotic behaviors of the wave functions $\mo_\pm,\ \mi_\pm$ as 
$z\ap 0,\,\infty$ are easily determined.  We note that
$$
\frac{k}{E}\ap \cases 1 & z \api;\\
-1 & z \ap 0.
\endcases
$$
To simplify the discussion, we assume, as in [14], that
$$
\inmp p\,dy=0.
$$
This assumption is not essential and could be dropped, though with a 
consequent increase in the
computational details.
Letting 
$$
\l=e^{P/2},
\qquad
P=\inxp p(y)\,dy,
$$
we easily deduce from the integral equations (2.3a,b), that
$$
\align
\lim_{z\api}\mo_{\pm} (x,z) =&\l^{\mp 1},
\qquad
\lim_{z\ap 0}\mo_{\pm} (x,z) =\l^{\pm 1},
\qquad z\in \Cal U_{\pm};\\
& \hskip 2.5in \tag2.4\\
\lim_{z\api}\mi_{\pm}(x,z) =&\l^{\pm 1},
\qquad
\lim_{z\ap 0}\mi_{\pm}(x,z)=\l^{\mp 1}
\qquad
z\in \Cal U_{\pm}.
\endalign
$$
Note that $\l$ is real when $p$ is real and has modulus 1 when $p$ is 
imaginary.

If $p$ and $q$ are in the Schwartz class, the reduced wave functions $\mo$ have asymptotic expansions in powers of
$z$. Substituting $\psi_+=\mo_+ e^{iEx}$ into (2.1) we get the following equation for $m=\mo_+$:
$$
m\pp +2iE m\p -(ikp+q)m=0,
$$
We seek a formal asymptotic expansion of $m$ in inverse powers of $z$; that is, we set
$$
m(x,z)\sim \sum_{j=0}^{\infty} m_j (x)z^{-j}
$$
and substitute this asymptotic expansion into the differential equation for $m$. At orders 0 and 1
we obtain
$$
im_0\p-\frac{i}{2}pm_0=0,
\qquad
m_0\pp+im_1\p-\frac{i p m_1}{2}-qm_0=0.
$$
>From these two equations we obtain
$$
m_0=e^{-P/2}=\l^{-1},
\qquad
q=\frac{p^2}{4}+\ddx \left(\frac{p}{2}+im_1 e^{P/2}\right)
\tag2.5
$$
  
As in the theory of the standard Schr\"odinger operator, the wave
functions 
satisfy the usual relations on $\Sigma$:
$$
\align
\phi_+(x,\xi)=&a(\xi)\psi_-(x,\xi)+b(\xi)\psi_+(x,\xi)\\
\tag2.6 \\
\phi_-(x,\xi)=&c(\xi)\psi_-(x,\xi)+d(\xi)\psi_+(x,\xi)
\endalign
$$
where $\phi_+(x,\xi)$ denotes the limiting value of $\phi_+(x,z)$ as 
$z\ap \xi$ from $\Cal U_+$,
etc. The coefficients in (2.6) relate the incoming and scattered
waves.
Thus, for example, from (2.6) it follows that for $\xi\in\Sigma$
$$
\phi_+(x,\xi)\sim \cases e^{-iEx} & x\ap -\infty; \\
a(\xi)e^{-iEx}+b(\xi)e^{iEx} & x\api.  \endcases
$$

The matrix of coefficients
$$
S=\lmat a & b \\ c & d \rmat
$$
is sometimes called the scattering matrix ([6]).
[?].)
The coefficients of $S$ may be computed in terms of Wronskians
of the wave functions, just as for the standard Schr\"odinger
equation. 
Since the Wronskian of solutions of (2.1) is independent of $x$, we
have $W(\phi_-,\phi_+)=W(e^{iEx},e^{-iEx})=2iE.$
It then follows from (2.6) that,
$$
\gather
a(\xi)=\frac{W(\phi_+,\psi_+)}{2iE} ,
\qquad b(\xi)=\frac{W(\phi_+,\psi_-)}{-2iE},\\
\hfill \tag2.7 \\
 c(\xi)=\frac{W(\phi_-,\psi_+)}{2iE} ,
\qquad d(\xi)=\frac{W(\phi_-,\psi_-)}{-2iE}. 
\endgather
$$
Moreover, 
$$
\align
W(\phi_+,\phi_-)=&-2iE=W(a\psi_-+b\psi_+,\,c\psi_-+d\psi_+)
=W(\psi_-,\psi_+)(ad-bc)\\
=&-2iE\det\,S;
\endalign
$$
hence $\det S=1$.

At a zero of $a(z)$ at  $z_j^+\in\Cal U_+$ the wave functions $\phi_+$ and
$\psi_+$ are linearly dependent, and we can write
$$
\phi_\pm (x,z_j^+)=c_j^+\psi_\pm (x,z_j^+).
$$ 
where $c_j^+$ are the associated coupling coefficients.
  Since $\phi_+$ decays exponentially as $x\ap
-\infty$
and $\psi_+$ decays exponentially as $x\api$, this common function
belongs
to $L_2$ hence constitutes an eigenfunction of (2.1), called the bound
state. Similar considerations hold on $\Cal U_-$. 
Thus, the bound states of the problem are given by zeroes of $a(z)$ in $\Cal U_+$ and $d(z)$ in
$\Cal U_-$. We denote these zeroes by $z_j^{\pm}$ respectively, and
the
 associated coupling coefficients by $c_j^{\pm}$.

We define generalized reflection coefficients
$$
r_+(\xi)=\frac{b(\xi)}{a(\xi)},
\qquad
r_-(\xi)=\frac{c(\xi)}{d(\xi)}, \qquad \xi\in\Sigma.\tag2.8
$$
When the zeroes of $a$ and $d$ are simple, the 
{\it scattering data} for the operator (2.1) is
the set
$$
\{r_\pm(\xi),\ \xi\in\Sigma;\ z_j^\pm\ ;\ c_j^\pm\ \}
$$

Under the mapping $z\mapsto -1/z$, $E\mapsto -E$ and $k\mapsto k$ so the differential equation
(2.1) is invariant. We denote this mapping by $s$, and write
$s.\psi(x,z)=\psi(x,-1/z)$.

\proclaim{Lemma 2.1} Under the mapping $s$ the wave functions
transform as $s.\psi_+=\psi_-$, $s.\mo_+=\mo_-$,
i.e. $\psi_+(x,-1/z)=\psi_-(x,z)$, etc. 
Moreover, for $z\in\Cal U_+$, $a(z)=d(-1/z)$; while $b(\xi)=c(-1/\xi)$
and $r_+(\xi)=r_-(-1/\xi)$ for $\xi\in\Sigma.$
\endproclaim
\demo{Proof} Under the mapping $z\mapsto -1/z$ the 
Volterra integral equations in (2.3a) for $\mo_+$ and $\mo_-$ 
are interchanged. Since the solutions are uniquely determined, 
they are the same. The result extends immediately to $\psi_\pm$
and $\phi_\pm$. The relations on the generalized reflection coefficients
then follow from their expressions in terms of the Wronskians.
\qed
\enddemo

When $q$ is real and $p$ is real or imaginary, the wave functions possess additional symmetries. Since $k=\sqrt{E^2-1}$, $k$ is 
real on $E^2>1$ and imaginary on the slit $-1<E<1$. Hence (2.1) is
invariant under Schwarz reflection across the slit if $p$ is real, and  under 
Schwarz reflection across $E^2>1$ if $p$ is imaginary. The slit lifts to the unit circle in the $z$-plane, while the rays $E^2>1$ lift to the real line in the $z$ plane. These observations lead to
the following result:

\proclaim {Lemma 2.2} When $p$ is real,  (2.1) is invariant under
Schwarz reflection across the unit circle and across the imaginary
axis in the $z$-plane; and the wave functions $\phi$ and $\psi$
satisfy the symmetry conditions $\phi_+(x,z)=\overline{\phi_-(x,1/\bar z)}=
\overline{\phi_+(x,-\bar z)}.$

When $p$ is imaginary, (2.1) is invariant
under Schwarz reflection across the real axis $\Im \,z=0$,
as well as under the reflection $z\rightarrow -\bar z^{-1}$; the wave functions
possess the corresponding symmetries: $\phi_+(x,z)=
\overline{\phi_-(x,\bar z)}=\overline{\phi_+(x,-1/\bar z)}.$
\endproclaim
\demo{Proof} Take the case of real $p$, and consider the wave function 
$\phi$. An easy calculation shows that, for $z\in\Cal U_+$, both
$\overline{\phi_- (x,1/\bar z)},$ and $\phi_+(x,z)$
are asymptotic to $e^{-i Ex}$ as $ x\apm$ and satisfy the same differential
equation.
Since solutions of (2.1) are uniquely determined by their asymptotics
as $x\apm$, they in fact coincide. The other cases are handled in the same way.
\qed \enddemo

\proclaim{Theorem 2.3} When $p$ is real or imaginary the scattering data
possess  additional symmetries: 

For $p$ real,  $a(\xi)=\overline{a(-\bar\xi)}=\overline{d(1/\bar\xi)}$ and $r_+(\xi)=\overline{r_+(-\bar \xi)}=\overline{r_-(1/\bar \xi)}$.
Moreover, if $z_j$  is a bound state, then so are
$-\overline{ z_j}$, $\overline{z_j}^{-1}$ and $-z_j^{-1}$. 
The four associated coupling coefficients are respectively $c_j$, 
$\overline{c_j}$, $\overline{c_j}$, and $c_j$.

When $p$ is imaginary, $a(\xi)=\overline{a(-1/\bar\xi)}=\overline{d(\bar\xi)}$  and $r_+(\xi)=\overline{r_+(-1/\bar \xi)}
=\overline{r_-(\bar \xi)}.$
 The bound states also appear in fours: 
$z_j,\, \overline{z_j},\, -z_j^{-1},\, -\overline{z_j}^{-1}$,
with corresponding coupling coefficients
$c_j,\ \overline{c_j}, \ c_j,$ $ \overline{c_j}.$

Finally,
$$
1-|r(\xi)|^2=1/|a(\xi)|^2 
\qquad
\cases |\xi|=1 & p\ \text{real}; \\ \Im\xi=0 & p\ \text{imaginary}.
\endcases
$$
\endproclaim

\demo{Proof} The lemma follows from Lemma 2.2 and the 
computation of the coefficients of $S$ in terms of Wronskians. 
>From $\det S=ad-bc=1$ we get $1-r_+r_-=1/ad$; and the last statement
above follows from the relationships between $a$ and $d$, $r_+$ and $r_-$ 
on $\Sigma$
when $p$ is real or imaginary. \qed \enddemo

The following result will be useful in our discussion of the vanishing lemma, in
\S3.

\proclaim{ Lemma 2.4} Let $p$ be real and consider the bound states
lying on $i\Bbb R\setminus\{0\}$. Then 
$$
 \frac12\inmp (2\cosh \,\a_j-p)\phi_j^2\,dx=
 \cases i\omega_j a\p (i\omega_j)c_j & 1<\omega_j<\infty; \\
-i\omega_jd\p (i\omega_j)c_j & 0<\omega_j<1,
\endcases \tag2.9a
$$
and
$$
\frac12 \inmp (2\cosh \,\a_j+p)\phi_j^2\,dx=
\cases -i\omega_ja\p (i\omega_j)c_j & -1<\omega_j<0; \\
i\omega_jd\p (i\omega_j)c_j & -\infty <\omega_j<-1,
\endcases \tag2.9b
$$
where $|\omega_j|=e^{\a_j}$ in both cases,
$c_j$ is the coupling coefficient
associated with the bound state, and $\phi_j=\phi_\pm(x,i\o_j)$
according as $\o_j\in\Cal U_\pm.$
\endproclaim
\demo{Proof} Differentiating (2.1) with respect to $z$ we have
$$
(D^2+E^2-V)\phi' =(V'-2EE' )\phi
\qquad
(D^2+E^2-V)\phi=0,
$$
where primes denote differentiation with respect to $z$, and $V=ikp+q$. Multiplying the first of these equations by $\phi$, the second by $\phi'$ and subtracting, we obtain
$$
D W(\phi,\phi')=(V'-2EE')\phi^2,
\qquad
W(\phi,\phi')=\phi D\phi' -\phi' D\phi .
$$
Now $\phi_j=\phi_+(x,i\omega)$ tends to zero exponentially as $x \ap \pm
\infty$, while $\phi_j^\prime$ tends to zero exponentially as $x\ap -\infty$.  Integrating the above expression over the real line we obtain
$$
\align
\lim_{x\api} W(\phi_j,\phi_j^\prime)=&\frac{i}{\omega_j}\inmp [p(y)
\sinh \,\a_j -2
\sinh\,\a_j
\cosh\,\a_j]\phi_j^2\, dy\\
=&\frac{i\sinh\,\a_j}{\omega_j}\inmp [p(y)\ - 2\cosh\,\a_j]\phi_j^2\, dy,
\endalign
$$
where $\omega_j=e^{\a_j}$ for the case $\omega_j>0$.
>From (2.7), $W(\phi_+,\psi_+)=2iEa(z)$ for
$z\in \Cal U_+$; differentiating this identity with
respect to $z$, and evaluating at a bound state $z=i\omega_j$ we obtain
$$
W(\phi_j',\psi_j)+W(\phi_j,\psi_j')=2iE_ja'(i\omega_j)=
-2\sinh\,\a_j a\p(i\omega_j),
$$
with
$\phi_j'=\phi_+'(x,i\omega_j),$ etc.

We now use the relation $\phi_j=c_j^+\psi_j$, in the
above result, and let $x\api$. The left side becomes 
$$
\frac{1}{c_j^+}W(\phi_j^\prime,\phi_j)+c_j^+W(\psi_j,\psi_j^{\prime}).
$$
The second term tends to zero, since 
$\psi_j$ and $\psi_j^\prime$ tend to zero exponentially as $x\api$. 
The identity (2.9a) now follows. The identity (2.9b) is obtained
in the same manner.\qed
\enddemo

As an immediate consequence of Lemma 2.4 we have:

\proclaim{Theorem 2.5} Let $p$ be real and $|p|<2$. Then all the bound states
$z_j=i\omega_j\in i\Bbb R\setminus\{0\}$  are simple, and  $C_j>0$, where
$$
C_j=\cases -\frac{A_j}{z_j} & 1<\omega_j<\infty; \\
\frac{A_j}{z_j} & -1<\omega_j<0, \endcases 
\qquad
A_j= -\frac{c_j}{a\p(z_j)}
\tag2.10
$$
\endproclaim
We now formulate the forward problem as a
 Riemann-Hilbert problem. Define the row vector $m$ by 
$$
m(x,z)=\cases \left(\psi (x,z), \frac{\phi (x,z)}{a(z)}\right)
e^{-i E x\sigma_3}
&  z \in \Cal U_+; \\
\left( \frac{\phi(x,z)}{d(z)},  \psi(x,z)\right)e^{i E x\sigma_3}
&  z \in \Cal U_-,
\endcases
$$
$$
=\cases \left( \mo_+ (x,z), \frac{\mi_+ (x,z)}{a(z)}\right) &  
z \in \Cal U_+; \\
\left( \frac{\mi_- (x,z)}{d(z)},  \mo_-(x,z)\right) &  z \in \Cal U_-.
\endcases
\tag2.11
$$
Here, as usual, $\sigma_3$ denotes the Pauli spin matrix
$$
\sigma_3=\lmat 1 & 0 \\ 0 & -1 \rmat.
$$

A simple computation shows that (2.6) are equivalent to the jump
relations
$$
m_+(x,\xi)\lmat 1 & -r_+e^{2iEx} \\ 0 & 1 \rmat
=m_-(x,\xi)\lmat 1 & 0 \\ -r_-e^{-2iEx} & 1  \rmat. 
$$
We write this as
$$
m_+(x,\xi)=m_-(x,\xi)v(x,\xi),
\qquad
\xi\in \Sigma
 \tag2.12
$$
where
$$
v(x,\xi)=\lmat 1 & r_+(\xi)e^{2i  E x} \\ -r_-(\xi)e^{-2i  E x} &
1-r_+r_- \rmat.
\tag2.13
$$

The row vector $m$ has prescribed asymptotic behavior as $x\api$ and
as $z$ tends to $0$ or $\infty$, namely:
$$
m(x,z) \ap (1,1), \quad \text{as}\ x\api \tag2.14
$$
and by (2.4)
$$
m(x,z) \ap \cases (\lambda,\lambda^{-1}) & z \ap 0; \\ 
(\lambda^{-1},\lambda) & z \api.
\endcases \tag2.15
$$

Moreover,
$$
\psi_+(x,i)=\mo_+(x,i)=\psi_-(x,i)=\mo_-(x,i). \tag2.16
$$
This identity follows by observing that $\psi_{\pm}(x,i)$ satisfy the same Volterra integral equation, viz.
$$
\psi_{\pm}(x,i)=1-\inxp (x-y)(q-p)\psi_\pm (y,i)\,dy.
$$
We also have $\psi_-(x,-i)=\psi_+(x,-i)$.

In the case of a reflectionless potential, 
$r_+(\xi)\equiv 0$, and $v(x,\xi)=I$ everywhere on $\Sigma$.

\proclaim{Theorem 2.6}  The inequality $|r(\xi)|^2<1$ 
holds on the entire real line in the
case $p$ imaginary, and everywhere on the unit circle except at 
$\pm i$ in the case $p$ real. When $p$ is real we have generically,
 (that is, except in the case of
reflectionless potentials) $r_+(\pm i)=r_-(\pm i)=-1$ and
$$
v(x,\pm i)=\lmat 1 & -1 \\ 1 & 0 \rmat.
$$
In all cases,  $r_\pm(0)=0,$ and $v(x,0)=I$.

\endproclaim
\demo{Proof} By Theorem 2.3 we have $|a|^2-|b|^2=1$ on either the real line 
or the unit circle, according as $p$ is imaginary or real; hence
$a$ cannot vanish on the corresponding portion of $\Sigma$.
 Writing this identity as
 $1-|r(\xi)|^2=|a(\xi)|^{-2}$, 
we see that 
$|r(\xi)|<1$  whenever 
$$
\frac{1}{a(\xi)}=\frac{2iE}{W(\phi_+,\psi_+)}\ne 0.
$$
We see that this holds whenever $E\ne 0$, hence everywhere except at
$\xi=\pm i$.
 
Writing the first equation in (2.6) in the form
$$
\frac{\phi_+(x,\xi)}{a(\xi)}=\psi_-(x,i)+r_+(\xi)\psi_+(x,\xi),
$$
and noting that $1/a(i)=0$ (note that, since we have already observed
that $a(i)\ne 0$ for real $p$,  $W(\phi_+(x,i),\psi_+(x,i))\ne 0$) 
we see that $r_+(i)=0$.  
A similar argument shows that $r_-(\pm i)=0$. The expression for
$v(x,\pm i)$ above follows from the fact that $E(\pm i)=0$.

 We write 
$r_+(\xi)=-W(\phi_+,\psi_-)/W(\phi_+,\psi_+).$
Then
$$
W(\phi_+,\psi_+)=W(e^{-iEx}\mi_+,e^{iEx}\mo_+)
=-2iE\mo_+\mi_++W(\mi_+,\mo_+).
$$
As $\xi\ap 0$, $\mo_+\ap \l$, and $\mi_+\ap \l^{-1}$, so that
$$
\lim_{\xi\ap 0}W(\mi_+,\mo_+)=W(\l^{-1},\l)=-2\ddx\log \l=p(x)
$$
A similar calculation shows that
$$
\lim_{\xi\ap 0}W(\phi_+,\psi_-)=\lim_{\xi\ap 0}e^{-2iEx}W(\mi_+,\mo_-)
=0
$$
since $\mi_+$ and $\mi_-$ both tend to $\l^{-1}$ as $\xi\ap 0$.
Noting that $E(\xi)\api$ as $\xi\ap 0$, we see that 
$r_+(\xi)\ap 0$ as $\xi\ap 0$. A similar argument shows
that $r_-(0)=0$.\qed
\enddemo

Since $1/a(i)=1/d(i)=0$ we have, in the generic case, 
$$
m_+(x,i)=(\mo_+(x,i),\, 0),
\qquad
m_-(x,i)=(0,\, \mo_-(x,i));
\tag2.17
$$

\proclaim{Lemma 2.7} The row vector $m$ possesses the symmetries
$$
s.m=m(x,-1/z)=m(x,z)R, \qquad R=\lmat 0 & 1 \\ 1 & 0 \rmat;
$$
and $v(x,\xi)=Rv^{-1}(x,-1/\xi)R.$ When $q$ is real, 
$m(x,z)=\overline{m(x,1/\bar z)}R$
if $p$ is real; and $m(x,z)=\overline{m(x,\bar z)}R$ when $p$ is
imaginary.
\endproclaim

The row vector $m$ has poles at the zeroes of $a$ and $d$. At a simple
pole of $m$, there is a
triangular matrix $v_j(z)$ such that
$$
\rho_j(x,z)=m e^{iEx\sigma_3}v_j(z) e^{-iEx\sigma_3},
$$
is regular at $z=z_j$ [1]. In fact, from the bound state relation $\phi_\pm(x,z_j^\pm)=
c_j^\pm\psi_\pm(x,z_j^\pm)$ we find this to be true if we take
$$
v_j=\lmat 1 & A_j(z-z_j^+)^{-1} \\ 0 & 1 \rmat, \qquad z_j^+\in\Cal U_+
\qquad
A_j=-\frac{c_j^+}{a'(z_j^+)};  \tag2.18a 
$$
$$
v_j=\lmat 1 & 0\\ D_j(z-z_j^-)^{-1}  & 1 \rmat, \qquad z_j^-\in\Cal U_-
\qquad
D_j=-\frac{c_j^-}{a'(z_j^-)}. \tag2.18b
$$

>From the symmetry $a(z)=d(-1/z)$, it follows that if $m$ has a pole at
$z_j\in\Cal U_+$, with coupling coefficient $c_j$, then it also has
one at 
$-1/z_j\in \Cal U_-$, with the same coupling coefficient. In fact, from $\phi_+(x,z_j)=c_j\psi_+(x,z_j)$ and the $s$-symmetry we find
$\phi_-(x,-1/z_j)=c_j\psi_-(x,-1/z_j)$.
>From the identity $a^\prime(z)=-z^{-2}d^\prime(-1/z)$, we find
$$
\frac{D_j}{(-1/z_j)}=-\frac{c_j}{z_ja\p(z_j)}
=\frac{A_j}{z_j} \tag2.19
$$

>From (2.18) we deduce that $m_1$ is regular in $\Cal U_+$ while
$m_2$ is regular in $\Cal U_-$, and that
$$
\text{Res}\ m(x,z)\Big|_{z=z_j}
=\cases (0, -A_jm_1(x,z_j)e^{2ixE_j}), & z_j\in\Cal U_+ ;\\
(-D_jm_2(x,z_j)e^{-2ixE_j},0), & z_j\in\Cal U_-, \endcases \tag2.20
$$
where $2E_j=(z_j+1/z_j)$.

It is easily seen that both components of the row vector $\rho e^{iEx\sigma_3}$ satisfy the differential equation (2.1). Moreover, $\rho_j$ satisfies (2.14), (2.15). Now choose a small circle $\Cal C_j$ around $z_j$, containing no other bound states
and not intersecting $\Sigma$, and define a new vector $\widetilde m$ by
$$
\widetilde m(x,z) =\cases  \rho_j(x,z)  & z \ \text{inside} \  \Cal C_j ;\\
        m(x,z) & z \ \text{outside}\ \Cal C_j .\endcases 
$$
Then the jump of $\widetilde m$ across the contour $\Cal C_j$ is
$$
\widetilde m_+=\widetilde m_-v_j(x,z),
\qquad
v_j(x,z)=e^{iE_jx\sigma_3}v_j(z)e^{-iE_jx\sigma_3}, \tag2.21
$$
where the limits $\widetilde m_{\pm}$ are taken relative to a
counterclockwise orientation of $\Cal C_j$ and $E_j=E(z_j)$. If there are only a finite number of simple poles, we can modify the original
contour by adding a small circle around each pole, and redefining $m$ accordingly,
so that the original Riemann-Hilbert problem is replaced by a modified one
on the extended contour.

\proclaim{Theorem 2.8} Let $p,\ q\in \Cal S$, and let $m$ be the reduced
vector valued wave function defined in (2.11). 
Then:
\roster
\item $m$ satisfies the Riemann-Hilbert problem
(2.12, 2.21) with $v$ and $v_j$ given in (2.13), (2.18);
\item  $m$ satisfies the asymptotic conditions (2.14), 
(2.15), and the
conditions (2.16) at $z=i$ together with corresponding 
conditions at $z=-i$;
\item in the generic case, $m$ satisfies the constraints (2.17), plus
similar constraints at $\xi=-i$;
\item  m satisfies the symmetries of Lemma 2.7.
\endroster
\endproclaim

\head 3. Two Vanishing Lemmas \endhead

A general approach to the solution of Riemann-Hilbert problems such
as (2.8) is given in [3]. The inverse problem is reduced to the
inversion of a linear operator of the form $I+T_1+T_2$ where
$T_1$ is of small norm and $T_2$ is compact. Once this reduction has
been obtained, the Fredholm alternative applies, and the proof of
existence and uniqueness is reduced to proving a uniqueness theorem
for the Riemann-Hilbert problem. Such a uniqueness theorem in the context of Riemann-Hilbert problems is sometimes called a
``Vanishing Lemma'' [3].
We prove the vanishing lemma for our problem in this section. 
The remaining details
of the inverse problem are the same as, but somewhat simpler than,
those given in Part II of
[3]. The inverse problem treated in that book, the Riemann-Hilbert
problem for an $n^{th}$ order ordinary differential operator, is
complicated  by the intersection of rays at the origin.

The proof of the vanishing lemma is based on the symmetries on the
reflection coefficients given in Theorem 2.3. For the purposes of the vanishing lemma, $x$ appears only as a parameter, so we suppress
it for the discussion.
We first prove a vanishing lemma for the Riemann-Hilbert problem that arises in the case when $p$ is imaginary and  there are no bound states. 

\proclaim{Theorem 3.1} Let $m$ be a piecewise analytic function of
$z$ in $\Im \,z\ne 0$, with boundary values $m_\pm\in L_2(\Bbb R)$; and let $m$ satisfy the Riemann-Hilbert problem
$$
m_+(\xi)=m_-(\xi)v(\xi), 
\qquad
\xi\in\Bbb R;
\qquad
m=O(z^{-1})
\qquad
\text{as} \ \ z\api,
$$
where
$$
v(\xi)=\lmat 1 & r(\xi) \\ -\overline{r(\xi)} & 1-|r(\xi)|^2 \rmat. \tag3.1
$$
Assume $|r|<1\ a.e.$ Then $m$ vanishes identically.
\endproclaim

\demo{Proof} We define
$$
F=\cases \frac{m(z)\cdot m^\dagger(\bar z) }{2} & \quad \Im\,z>0; \\
-\frac{m(z)\cdot m^\dagger(\bar z) }{2} & \quad \Im\,z<0,
\endcases
$$
where
$$
m^\dagger=\lmat \overline{m_1} \\ \overline{m_2} \rmat.
$$
Thus, $m(z)\cdot m^\dagger(\bar z)=m_1(z)\overline{m_1(\bar z)}+m_2(z)\overline{m_2(\bar z)}$. 

By the Schwarz reflection principle, $F$ is sectionally analytic in $\Im z\ne 0$. Its boundary values on the real line are
$$
F_+(\xi)=\frac{m_+(\xi)\cdot m_-^\dagger(\xi)}{2}=\frac{m_-(\xi)v(\xi) m_-^\dagger(\xi)}{2};
$$
and
$$
F_-(\xi)=-\frac{m_-(\xi)\cdot m_+^\dagger(\xi)}{2}=-\frac{m_-(\xi)v^\ast(\xi)m_-^\dagger(\xi)}{2}.
$$

Therefore, the jump of $F$ across the real line is 
$$
[F]=F_+-F_-=m_-\left(\frac{v+v^\ast}{2}\right)m_-^\dagger
=|m_-^1(\xi)|^2+(1-|r(\xi)|^2)|m_-^2(\xi)|^2.
$$
Since $F$ has no
poles and is $O(z^{-2})$ as $z\api$, we can integrate it around a 
circle of radius $R_a$ centered at the origin, and let $R_a\api$ to
obtain
$$
\inmp [F]\,d\xi=\inmp |m_-^1(\xi)|^2+(1-|r(\xi)|^2)|m_-^2(\xi)|^2\,d\xi= 0.
$$

Since $|r|<1\ a.e.$ the boundary values $m_-(\xi)$ and $m_+(\xi)=m_-(\xi)v(\xi)$
vanish {\it a.e.} on the real line. Since $m\ap 0$ as $z\api$ and has no other singularities,
$m$ must vanish identically. \qed

\enddemo

We now prove a vanishing lemma for the case corresponding to real $p$.
The operator (2.1) is self-adjoint only
for $z$ on the imaginary axis or the unit circle, so in this case we restrict the scattering data
to lie on the union of the imaginary axis and the unit circle. 
Note also that the data and wave functions are invariant under Schwarz reflection in the unit circle and in the imaginary axis.

\proclaim{Theorem 3.2} Let $m(z)$ be a piecewise meromorphic
function of $z$ in $|z|\ne 1$ satisfying the Riemann-Hilbert problem
$m_+(\xi)=m_-(\xi)v(\xi)$ on $|\xi|=1$, where $v$ has the form (3.1) on the unit circle.
Assume that $|r(\xi)|<1 \ a.e.$ on the unit circle, and that
\roster
\item the poles of $m$ are simple and lie on the imaginary axis; 
\item  the residue matrices at $z_j=i\o_j$ and $z_{-j}=i/\o_j$ are
given by (2.18a) and (2.18b) respectively, where $A_j$ and $D_j$ satisfy (2.19)
and $A_j/z_j<0$.
\item both components of $m$ are $O(z^{-1})$ as $z\rightarrow \infty$;
\endroster
Then $m$ vanishes identically.
\endproclaim
\demo{Proof} We proceed in three steps. 

{\it Step 1.} We first prove the theorem in the case there are no bound states. 
 Define the function
$$
F(z)=\cases -\frac{m(z)\cdot m^\dagger (1/\bar z)}{2z} & |z|>1; \\
\frac{m(z)\cdot m^\dagger (1/\bar z)}{2z} & |z|<1.
\endcases
$$
By the Schwarz reflection principle, $F$ is sectionally analytic in $|z| \ne 1$; and also 
$$
F(z) = \cases  O(z^{-2})  & z\api; \\
O(1) & z \ap 0. \endcases  \tag3.2
$$

We orient the unit circle in the counterclockwise direction, and denote the
limiting values of $F$ from the interior and exterior of the unit
circle 
by $F_+$, $F_-$ respectively.  
On $|\xi|=1$,
$$
F_+(\xi)=\frac{m_-(\xi)\cdot m_+^\dagger(\xi)}{2\xi }
=\frac{m_-(\xi)v^\ast (\xi)m_-^\dagger(\xi)}{2\xi };
$$
and 
$$
F_-(\xi)=-\frac{m_+(\xi)\cdot m_-^\dagger(\xi)}{2\xi }
=-\frac{m_-(\xi)v(\xi)m_-^\dagger(\xi)}{2\xi }.
$$
(Note that $\xi=1/\bar \xi$ on the unit circle.)
The jump of $F$ across the unit circle is therefore
$$
\align
[F]=&m_-\left(\frac{v+v^\ast}{2\xi}\right)m_-^\dagger \\
=&\frac{|m_-^1(\xi)|^2+(1-|r(\xi)|^2)|m_-^2(\xi)|^2}{\xi}, 
\qquad |\xi|=1. \tag3.3
\endalign
$$

Let $\Gamma_{R_a}^-$ be the contour consisting of the unit circle 
traversed in the clockwise direction and the circle $|z|=R_a>1$ in
the 
counterclockwise direction; and let $\Gamma_{R_a}^+$ consist of the unit circle
traversed in the counterclockwise direction together with the circle
$|z|=1/R_a$ traversed in the clockwise direction. We let
$\Gamma_{R_a}=\Gamma_{R_a}^+ \cup \Gamma_{R_a}^-$. 
For all $R_a>1$ we get, 
by Cauchy's theorem,
$$
\frac{1}{2\pi i}\int_{\Gamma_{R_a}}F(z)\,dz=0.
$$
since $F$ has no poles. Letting $R_a\rightarrow \infty$ we obtain, in virtue of (3.2),
$$
\frac{1}{2\pi }\int_{|\xi|=1}|m_-^1(\xi)|^2+(1-|r(\xi)|^2)|m_-^2(\xi)|^2 \frac{d\xi}{i\xi}=0. 
$$
Now $d\xi/ i\xi$ is a 
positive measure on the unit circle oriented in the counterclockwise
direction,  and $|r_+(\xi)|<1 \ a.e.$
The boundary values $m_-$ and $m_+=m_-v$ must therefore vanish {\it a.e.} 
on the unit circle. Since $m$ has no singularities in the finite plane and 
vanishes at $z=\infty$, $m$ vanishes identically. This completes step 1.

{\it Step 2.}  We next prove the theorem when $m$ has simple poles at $i\o_j$ and $i/\o_j$, 
where $\o_j>1$ and
$1\le j \le N$ and there are no jumps across
the unit circle. We define
$$
F(z)=\cases -\frac{m(z)\cdot m^\dagger (-\bar z)}{2z}& |z|>1; \\
\frac{m(z)\cdot m^\dagger (-\bar z)}{2z} & |z|<1. \endcases
$$

We now calculate the residue of the first term at a pole $z_j=i\o_j$,
$\o_j>1$. 
Since $\o_j>1$, $m_1$ is regular at $z_j=i\o_j$,
and, by (2.20),
$$
\Res m_2(z)\Big|_{z=i\o_j}=-A_jm_1(i\o_j).
$$
Now observe that if the function $f(z)$ has a simple pole at $z_0$ with residue
$f_0$,
then the function $\overline{f(-\bar z)}$ has a pole at $-\bar z_0$ with residue
$-\overline{f_0}$. Therefore
$$
\Res \overline{m_2(-\bar z)}\Big|_{z=i\o_j}=\overline{A_jm_1(i\o_j)},
$$
and
$$
\Res \frac{m_1(z) \overline{m_2(-\bar z)} }{2z} \Big|_{z=i\o_j}=
\frac{\overline{A_j} } {2i\o_j} |m_1(i\o_j)|^2
=-\frac{A_j}{2z_j} |m_1(i\o_j)|^2
$$
(Note that, since $A_j/z_j<0$, $A_j$ is imaginary.) 

The residue of the second term is calculated in a similar manner, and we find that
$$
\Res F \Big|_{z=i\o_j}=\frac{A_j}{z_j} |m_1(i\o_j)|^2.
$$
A similar calculation yields
$$
\Res F \Big|_{z=i/\o_j}=\frac{D_j}{-1/z_j} |m_1(i\o_j)|^2=\frac{A_j}{z_j} |m_1(i\o_j)|^2.
$$
Here we have used (2.19).

We again integrate $F$ around the contour $\Gamma_{R_a}$ and let 
$R_a\api$. By the Cauchy residue theorem,
$$
\lim_{R_a\api}\frac{1}{2\pi i}\int_{\Gamma_{R_a}}F(z)\,dz=2\sum_{j}C_j|m_-^1(i\o_j)|^2=0.
$$
where $C_j=-A_j/z_j$.

Since $C_j\ge 0$ by assumption 2, each term in the above sum must vanish. Since $m$ has no singularities in the plane and tends to 0 at infinity, $m$ must vanish identically.
This completes step 2.

{\it Step 3}. When $m$ has both poles and jumps across the unit circle, we construct a gauge
transformation that removes the poles and reduces the problem to the case where there are
only jumps. We construct a matrix valued function of $z$ with the following properties:
\ss
i) $M$ is meromorphic in the extended complex plane with simple poles at $z_j$ and $M\ap A$
as $z\api$, where $A$ is an invertible matrix. 
\vskip 1mm
ii) $Mv_j$ is regular at $z_j$.
\vskip 1mm
iii) $M(\xi)$ is unitary on the unit circle.
\ss
Given such an $M$ we put $m=wM$, where $m$ is the null solution of the original problem.
Since $mv_j$ and $Mv_j$  are regular at $z_j$, it follows that $w$ is regular at $z_j$. 
Moreover, $w=O(z^{-1})$ at infinity. The jumps of $w$ on the unit circle are given by
$$
w_+=w_-MvM^{-1}=w_-MvM^\ast
$$
since $M$ is unitary on the unit circle. We now apply the argument of step 1 to the row vector
$w$, which has no poles in the complex plane,  with
$v$ replaced by $MvM^\ast$. Then
$$
M\frac{v+v^\ast}{2}M^\ast
$$
is also positive definite, and we may conclude that $w_\pm$ vanish on $|\xi|=1$.

It remains to construct the matrix $M$. Each of the rows of $M$ separately satisfy the
same vector Riemann-Hilbert problem as the row vector $m$, 
namely each row vector of $Mv_j$ is regular at $z_j$. So we need two row vector solutions of this problem, with distinct limits at infinity. We start by taking these limits to be given by $(1,0)$ and $(0,1)$ respectively, so that initially $M(\infty)=I$. These are 
finite dimensional, algebraic
problems, so that uniqueness implies existence. But we have already proved uniqueness of this problem
in Step 2. So there exists a unique $M$ satisfying i) and ii) with $M(\infty)=I$.

We next prove that
$$
M(z)M^\ast\left(\frac{1}{\bar z}\right)=M(0). \tag3.4
$$
By the symmetry $v_j^{-1}(z)=v^\ast(1/\bar z)$, we have, in the neighborhood
of $z=z_j$,
$$
M(z)M^{\ast}(1/\bar z)=M(z)v_j(z)v_j^\ast(1/\bar z) M^\ast (1/\bar z).
$$
By assumption the poles of $M$ are such that the row vectors of
$Mv_j$ are regular at $z=z_j$.  
 Now
$$
v_j^{\ast \, -1}\left(\frac{1}{\bar z}\right)=
\lmat 1 & 0 \\ \frac{-zz_jD_j}{z+1/z_j} & 1 \rmat;
$$
and this matrix has the same residue at $z=-1/z_j$ as the matrix $v_j^-(z)$
in (2.18b). So $v_j^\ast(1/\bar z) M^\ast (1/\bar z)$ is also
 regular in a neighborhood of
$z=z_j$. Since $M(z)M^\ast(1/\bar z)$ has no singularities in the finite $z$-plane and is bounded
at infinity, it is a constant matrix, by Louiville's theorem. 
Letting $z\ap 0$ we obtain (3.4).

Taking $z=i$ in (3.5), we see that $M(0)=M(i)M^\ast (i)$, and so is a positive definite matrix. Let
$A$ be the positive definite square root of $M(0)^{-1}$. Replacing $M(z)$ by
$AM(z)$ we obtain the required gauge transformation satisfying i)-iii) above.
\qed
\enddemo

\head 4. The Inverse Problem \endhead

In the formulation of the forward problem, the asymptotics of the row
vector $m$ at $z=0,\ \infty$ are given in terms of $\l(x)$
by (2.15). In the inverse problem, $\l$ is determined by requiring 
that the solution of the Riemann-Hilbert problem satisfy (2.15). In addition,
(2.16) must be satisfied, and this leads to an additional constraint on $\l$.
In the generic case, the constraint (2.17) must also be satisfied, leading
to yet another constraint on $\l$.
One must show that these various determinations of $\l$ are consistent.

Consider the matrix Riemann-Hilbert problem
$$
\align
H_+(x,\xi) =& H_-(x,\xi)v(x,\xi), \qquad \xi\in\Sigma 
\qquad
Hv_j\ \text{regular at}\ z=z_j \\
&H\rightarrow I \ \text{as}\ z\api.
\endalign
$$
Each row of $H$ satisfies the Riemann-Hilbert problem of the previous
section.
Hence the vanishing lemmas of the previous section imply that this
problem has a unique solution $H$ when $v$ satisfies the appropriate
symmetries. Since $\det v=\det v_j=1$, the scalar function $\det H$ has no jumps and
no singularities, hence is identically 1.

Put 
$$
m=(1,1)\Lambda H,
\qquad
\Lambda=\lmat \l^{-1} & 0 \\ 0 & \l \rmat. 
$$
Then $m$, being a linear combination of the rows of $H$, satisfies the
Riemann-Hilbert problem, and $m\sim (\l^{-1},\l)$ as $z\api$.
We determine $\lambda(x)$ by requiring that (2.15) be satisfied, i.e.
$m\sim  (\lambda,\lambda^{-1} ), \ z \ap 0.$ (By Theorem 2.6 we may assume that
$v(x,0)=I$, hence that $H_+(x,0)=H_-(x,0);$ hence we may write unambiguously
$H(x,0)$ for $H_\pm(x,0)$.) 
This leads to:
$$
\l =\l^{-1}H_{11}(x,0)+\l H_{21}(x,0),
\qquad
\l^{-1}=\l^{-1}H_{12}(x,0)+\l H_{22}(x,0);
$$
hence to two determinations for $\l$:
$$
\l^2=\frac{H_{11}(x,0)}{1-H_{21}(x,0)},
\qquad
\l^2=\frac{1-H_{12}(x,0)}{H_{22}(x,0)}.
 \tag4.1
$$

Moreover,  (2.16) implies, 
$\l^{-1}H_{11}^+(x,i)+\l H_{21}^+(x,i)=\l^{-1}H_{12}^-(x,i)+\l H_{22}^-(x,i),$
so
$$
\l^2=\frac{H^+_{11}(x,i)-H^-_{12}(x,i)}{H^-_{22}(x,i)-H^+_{21}(x,i)}. \tag4.2
$$
In the case of reflectionless potentials, or when $v=I$ on the unit circle,
we may take $H_+(x,i)=H_-(x,i)$ in (4.2).
In addition, in the generic case, the equations in (2.17) imply
$$
\l^{-1}H_{12}^+(x,i)+\l H_{22}^+(x,i)=0,
\qquad
\l^{-1}H_{11}^-(x,i)+\l H_{21}^-(x,i)=0;
$$
hence
$$
\l^2= - \frac{H_{12}^+(x,i)}{H_{22}^+(x,i)}=
-\frac{H_{11}^-(x,i)}{H_{21}^-(x,i)}. \tag4.3
$$
A similar determination of $\l$ is given at $-i$.

We must show the equivalence of these various determinations of $\l$.
We first prove
\proclaim{Lemma 4.1} Assume the Riemann-Hilbert
problem above has a unique solution. Then
$$
RH(x,0)R=H^{-1}(x,0), \tag4.4
$$
$$
H_+(x,i)=H(x,0)RH_-(x,i)R.
 \tag4.5
$$
\endproclaim
\demo{Proof}Let $\widetilde H(x,z)=RH(x,-1/z)R.$
By the $s$-symmetry of the data, $\widetilde H$ and $H$ satisfy the same Riemann-Hilbert problem.
Since the solution of that problem is uniquely
determined by its behavior at $\infty$, we must have
$$
\widetilde H(x,z)=RH(x,-1/z)R=MH(x,z), \tag4.6
$$
where $M=M(x)$ is a matrix independent of $z$, and $\det M(x)=1$.
Letting $z\ap 0,\ \infty$ we find $RH(x,\infty)R=I=MH(x,0)$
and $RH(x,0)R=M$, hence (4.4).
Writing (4.6) as $H(x,z)=H(x,0)RH(x,-1/z)R$ and
letting $z\ap i$ from $\Cal U_+$, we obtain (4.5).\qed
\enddemo

We are now ready to show the equivalence of (4.1), (4.2), and, 
in the generic case, (4.3).
The two expressions in (4.1) are equivalent provided
$$
H_{12}(x,0)+H_{21}(x,0)=1-\det H(x,0).
$$
>From the fact that $\det H=1$, 
we see that we must show that $H_{12}(x,0)+H_{21}(x,0)=0$; but 
this follows from the identity (4.4). 

The equivalence of (4.1) and (4.2) reduces to the identity
$$
\align
H_{11}(x,0)&(H_{22}^-(x,i)-H_{21}^+(x,i))=\\
&H_{11}^+(x,i)-H_{12}^-(x,i)-H_{21}(x,0)(H_{11}^+(x,i)-H_{12}^-(x,i)).
\tag4.7
\endalign
$$
In the case of reflectionless
potentials, $v(x,\xi)=I$, and  $H_+(x,i)=H_-(x,i)$. In (4.7), replace
$H_-(x,i)$ by $H_+(x,i)$, $H_{21}(x,0)$ by $-H_{12}(x,0)$; one then obtains
the same expression as that obtained by operating on the column vector
$(1,-1)^\dagger$ with both sides of the matrix equation (4.5).

In the generic case we
write (4.5) in the form $H(x,0)RH_+(x,i)=H_+(x,i)Rv(i)$ 
and use the expression for $v(i)$ given in Lemma 2.6. We obtain four
equations, one of which is
$$
H_{11}(x,0)H_{22}^+(x,i)+H_{12}(x,0)H_{12}^+(x,i)= - H^+_{12}(x,i).  
$$
This may be used to show the equivalence of 
(4.1), (4.2). The proofs of the equivalence of (4.3) with (4.1) in the generic
case proceeds along similar lines, and is left to the reader.

\proclaim{Theorem 4.2} Let $m(x,z)=(1,1)\Lambda H(x,z)$. Then $m(x,-1/z)=m(x,z)R.$
\endproclaim
\demo{Proof} Since $\l$ has been chosen so that $m(x,0)=(\l,\l^{-1})$
we have $(\l,\l^{-1})=$ $(1,1)\Lambda H(x,0);$
so
$$
\align
m(x,z)=&(1,1)\Lambda H(x,0)H^{-1}(x,0)H(x,z)=(\l,\l^{-1})\widetilde H(x,z)\\
=&(\l,\l^{-1})R H(x,-1/z)R\\
=& (\l^{-1},\l)H(x,-1/z)R=(1,1)\Lambda H(x,-1/z)R\\
=&m(x,-1/z)R. \qed
\endalign
$$
\enddemo

\proclaim{Theorem 4.3}  Let the scattering
data $v$ satisfy the symmetry $\overline{v(\bar \xi)}$
$=Rv^{-1}(\xi)R.$
Then $|\l|=1.$
\endproclaim
\demo{Proof}
Define $\widetilde H(x,z)=\overline{H(x,\bar z)}.$
A short calculation shows that
$$
\align
\widetilde H_+(x,\xi)=&\lim_{z\ap \xi^+} \overline{H(x,\bar z)}=\overline {H_-(x,\bar \xi)}\\
=&\overline{H_+(x,\bar \xi)}\overline{v^{-1}(\bar \xi)}=
\overline{H_+(x,\bar \xi)}Rv(\xi)R\\
=&\widetilde H_-(x,\xi)Rv(\xi)R.
\endalign
$$
Since the solution of the Riemann-Hilbert problem is uniquely determined by
its limit as $z\api$ and since
 both $H$ and $\widetilde H$ tend to $I$ as $z \api$ we have 
$\widetilde H(x,z)$ $=RH(x,z)R.$
Letting $z\ap 0+$, and recalling that $H_+(x,0)=H_-(x,0)$,
we obtain $H(x,0)=R\overline{H(x,0)}R$, hence
$H_{11}(x,0)=\overline{H_{22}(x,0)}$ and $H_{12}(x,0)=\overline{H_{21}(x,0)}$.
The two expressions for $\l^2$ in (4.1) then show that $\l^2=
\overline{\l^{-2}}$.\qed
\enddemo

For real $p$, we find that $H_{\pm}(x,z)=\overline{H_{\pm}(x,-\bar z)},$
hence $ H_{\pm}(x,i)=\overline{H_{\pm}(x,i)}.$
>From (4.2) we see that $\l^2$ is real. We still need to show that this quantity is positive.
        
Finally, as in [3] one may prove that $H\rightarrow I$ as $x\api$,
so that (2.14) is satisfied.

\head  5. The Initial Value Problems \endhead

We want to solve the initial value problems for (1.2) and (1.3) by the
inverse scattering method. We begin by determining the associated
evolution
of the scattering data.

The evolution of the scattering data is obtained by standard asymptotic arguments.
>From [14] the flows (1.2), (1.3) are obtained from the Lax pair
$$
L=D^2-u,
\qquad
u=ikp+q;
\qquad
P=-i\frac{\e p_x}{4}+\e(\frac{ip}{2}+k)D
$$
We take $\e=2i$ for the flow (1.2) and $\e=2$ for the flow (1.3) [14].
The flows are obtained from the Lax equations
$$
\dot L=[P,L]
$$
when the commutator is evaluated on the submanifold of wave functions of $L$;
that is, on functions satisfying $L\psi+k^2\psi=0$.

 When the Lax equations are satisfied,
$$
(\frac{\partial}{\partial t}-P)(D^2-u+E^2)\phi=
(D^2-u+E^2)(\frac{\partial}{\partial t}-P)\phi=0
$$
so 
$$
(\frac{\partial}{\partial t}-P)\phi
$$
satisfies (2.1) whenever $\phi$ does. 

Since $p\ap 0$ to as $x \ap \pm \infty$,
$$
\frac{\partial}{\partial t}-P\sim \frac{\partial}{\partial t}-\e k D,
\quad \text{as}\ x\ap
 \pm \infty.
$$
so
$$
\left(\frac{\partial}{\partial t}-P\right)\phi \sim 
-\e k D e^{-iEx}
\sim
i\e\,kEe^{-iEx},
\qquad
\text{as}\ x\apm.
$$
Since the wave functions are uniquely determined by their asymptotic behavior
as $x\rightarrow -\infty$,
$$
(\frac{\partial}{\partial t}-P)\phi=i\e kE\phi.
$$

On the other hand, $\phi_+=a\psi_- +b\psi_+$, so on $\Sigma$,
$$
\align
(\frac{\partial}{\partial t}-P)\phi_+=&(\frac{\partial}{\partial t}-P) (a\psi_- +b\psi_+)\\
\sim & (\dot a+ikE \e a)e^{-iEx}+(\dot b-ikE\e b)e^{iEx}
\qquad x\api .
\endalign
$$
Therefore,
$$
(\frac{\partial}{\partial t}-P)\phi_+
= (\dot a+ikE\e a)\psi_- +(\dot b-ikE\e b)\psi_+
=i\e kE(a\psi_- +b\psi_+)
$$
and
$$
\dot a=0, \qquad  \dot b=2i\e kEb.
$$
Similarly  $\dot c=-2i\e kE c$. 
The evolution of the coupling coefficients for the bound states, defined by,
$$
\phi(x,t,z_j^{\pm})=c_j^{\pm}(t)\psi(x,t,z_j^{\pm})
$$
is derived by similar arguments. We have shown

\proclaim{Theorem 5.1} Under the flows (1.2) or (1.3) the evolution of the scattering data is given by
$$
r_\pm(x,t,\xi)=r_\pm(\xi)e^{\pm 2iE(x+\e kt)};
\qquad
c_j^{\pm}(t)=c_j^{\pm}(0)e^{ 2i\e k_j^{\pm}E_j^{\pm} t}, \tag5.1
$$
where $k_j^{\pm}$ and $E_j^{\pm}$ are the values of $k$ and $E$ at $z_j^{\pm}$.
\endproclaim

Global
existence theorems for each of the flows (1.2), (1.3) are proved
by showing that the scattering problem can be inverted at all later times, 
hence, by showing that the hypotheses on the scattering data in the
two vanishing lemmas are invariant under the evolution of the scattering data (5.1).
Now, the support of $r_\pm$ and the location of the bound states is 
invariant under the evolution of the scattering data, and $E$ is
real on $\Sigma$. Therefore, the only remaining constraints to be verified 
are that $\e k(\xi)$ is real 
on the support of $r_\pm(\xi)$
and that $i\e k_j^\pm E_j^\pm$ is real on the bound states $z_j^\pm$
in the case of real $p$.

We restrict the support of $r_\pm$ to lie on the unit circle in the case
$p$ real, and on the real line for the case $p$ imaginary. Since $k$ 
is real on the real line, and imaginary on the unit circle, and
 since $\e=2i$ in the Boussinesq case and $\e =2$ for imaginary $p$,
$\e k(\xi)$ is then always real on the support of $r_\pm$.

The condition that $ k_j^\pm E_j^\pm$ be real on the bound states
is met
for $z_j=i\o_j$ since in that case, $E_j=i\sinh \o_j,$ and
$k_j=i\cosh \o_j.$

The following theorem now follows as a consequence of the invariance of the scattering data
and the solvability of the inverse scattering problem.

\proclaim{Theorem 5.2} The initial value problem  (1.2) is solvable for $-\infty<t<\infty$
when, at some finite time, the support of the reflection coefficients $r_{\pm}(\xi)$ is contained in the unit circle, and the bound states are restricted to the imaginary axis and satisfy (2.10).

The initial value problem (1.3) is globally solvable if at some finite time the support of the reflection coefficients is contained in the real line, and their are no bound states. 
\endproclaim
\vskip .3in
\centerline{\epsfxsize=8cm \epsfbox{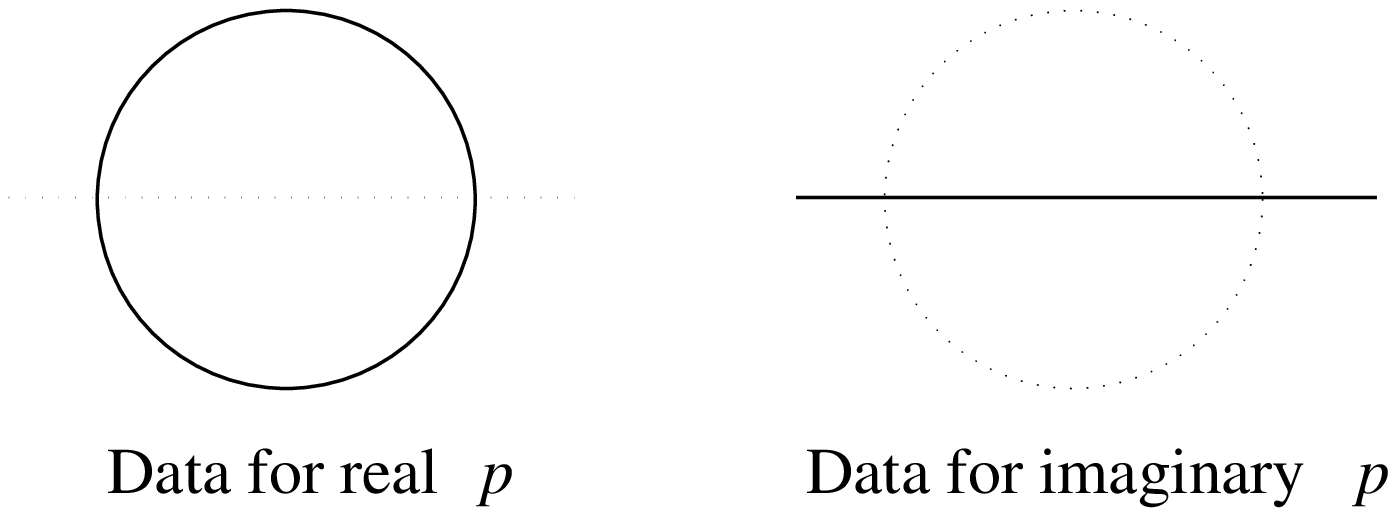}}
\vskip .3in

Global solutions of (1.3) do not exist when there are bound states for the case $p$ imaginary.
This is illustrated by the example in 
\S6 where we construct a soliton for (1.3).
In order to prove the vanishing lemma, we need a positivity condition,  as that of the expression
on the right side of (3.5). In the case of real $p$ we used the fact that  $m_1^+(x,i\omega)$ was real in (3.4).
In order to get real wave functions we need the coefficients in (1.1) 
to be real.
This means that $E$ is real or imaginary and $ikp$ is real, hence that
$k$ is real.

\head 6. Construction of Solitons \endhead

We construct simple solitary waves by solving the Riemann-Hilbert problem in the case of
reflectionless potentials, that is, when $v$ is the identity matrix and there are only discrete scattering data in
the problem.
When $p$ is real the bound state eigenvalues are symmetric with
respect to the imaginary axis and the unit circle, by Lemma 2.4. 
In the simplest case we may consider a pair of eigenvalues $i\o$ and $i/\o$, with $\o>1$. 

Our problem is
$$
\phi_+(x,z)=a(z)\psi_-(x,z),
\qquad
\phi_+(x,i\o)=c\psi_+(x,i\o) \tag6.1
$$
where $c$ is the coupling constant and
$$
a(z)=\frac{z-i\o}{z-i/\o}
\qquad
\psi_-(x,z)=\overline{\psi_+(x,\bar z^{-1})}. \tag6.2
$$

We take $\psi_+=\mo_+e^{ixE}$ and assume $\mo_+$ is a meromorphic function in the extended $z$-plane with a single pole at $i/\o$:
$$
\mo_+=\l^{-1}+\frac{w(x)}{z-i/\o}.
$$
>From (2.4),
$$
\lim_{z\ap 0}\mo_+(x,z)=\l
$$
hence
$$
\l-\l^{-1}=i\o w. \tag6.3
$$

Invariance of the wave function under Schwarz reflection in the imaginary axis, {\it viz.} 
$$
\overline{\psi_+(x,-\bar z)}=\psi_+(x,z),
$$
leads to the constraint
$$
\overline{\l^{-1}}-\frac{\bar w(x)}{z-i/\o}=\l^{-1}+\frac{w(x)}{z-i/\o}
$$
for all $z$. Hence we must have
$$
\bar \lambda=\lambda,
\qquad
\bar w=-w.
$$

Setting
$$
W=\frac{w}{2i\b},
\qquad
\l=e^{P/2},
\qquad
\o=e^{\a},
\qquad
\b=-iE(i\o)=\sinh\,\a,
$$
we obtain from equations (6.1-6.3) the relations
$$
-\o^2W=c[\l^{-1}+W]e^{-2\b x},
\qquad
W=-\o\b \sinh \frac{P}{2} .
$$
Solving for $P$, we obtain
$$
e^P=1+\frac{2c }{\b \o^3\exp\,(2\b x)+c\o\b } \tag6.4
$$
Then, from (2.5),
$$
\align
q=&\frac{p^2}{4}+\ddx \left(\frac{p}{2}-2\b W e^{P/2}\right)\\
= & \frac{p^2}{4}+\ddx
\left[\frac{p}{2}+\frac{2\b c }{\o^2\exp\,(2\b x)+c}\right]   \tag6.5
\endalign
$$

By Theorem 5.1 the time evolution of the coupling constant $c$ is
$$
c(t)=c_0e^{2t\sinh 2\a}.
$$

The eigenvalues for imaginary $p$ appear in fours, with the symmetries
given in  Theorem 2.3. We construct a soliton corresponding to four bound
states
with purely imaginary $z_0=i\o$, hence
$$
i\o,\quad i/\o, \quad -i/\o, \quad -i\o.
$$
Then
$$
a(z)=\frac{z-i\o}{z+i\o}\frac{z+i/\o}{z-i/\o},
$$
and
$$
\phi_+(x,z)=a(z)\psi_-(x,z), 
\qquad
\psi_-(x,z)=\overline{\psi_+(x,\bar z)}. \tag6.6
$$
>From Theorem 2.3
$$
\phi_+(x,i\o)=c \psi_+(x,i\o),
\qquad
\phi_+(x,-i/\o)=\bar c\psi_+(x,-i/\o) \tag6.7
$$
where $c$ is the coupling coefficient.

By Theorem 5.1, the time evolution of $c$ is given by
$$
c(t)=c(0)e^{2i\e kEt}=c(0)e^{-2i(\sinh\,2\a)t}.
$$

Now $\psi_+(x,z)=\mo_+(x,z)e^{iEx}$. We look for a wave function of
the form
$$
\mo_+(x,z)=\l^{-1}+\frac{w_1(x)}{z+i\o}+\frac{w_2(x)}{z-i/\o}.
$$
 The symmetry
$$
\mo_+(x,z)=\overline{\mo_+(x,-\bar z^{-1})}
$$
implies that
$$
\mo_+(x,z)=\bar \l^{-1}-\frac{z \bar w_1}{1+i\o z}-
\frac{z \bar w_2}{1-iz/\o }.
$$
The requirement that $\mo_+(x,0)=\l(x)$ implies that
$\l=\bar \l^{-1}$, hence that $|\l|=1$.

We then have
$$
\l-\l^{-1}=\left(\frac{w_1}{z+i\o}+\frac{z \bar w_2}{1-iz/\o}\right)
+\left(\frac{w_2}{z-i/\o}+\frac{z \bar w_1}{1+i\o z}\right).\tag6.8
$$
Since the right side is independent of $z$, 
$$
\bar w_1+\o^2w_2=0;
$$
and
$$
\mo_+(x,z)=\l^{-1}+\frac{w_1(x)}{z+i\o}+\frac{\bar w_1(x)}{i\o(1+i\o z)}.
$$
Equations (6.6) and (6.7) now lead to an equation for $w=w_1$:
$$
\bar w=\k\left(\l^{-1}+\frac{w}{2i\o}-
\frac{\bar w}{2i\o^2\,\sinh\,\alpha}\right)
e^{-2x\sinh\,\alpha } \tag6.9
$$
where
$$
e^{\alpha}=\o,
\qquad
\kappa = 2i\o c \tanh \,\alpha.
$$

Taking complex conjugates of (6.9) we obtain the following system of equations
for $w$ and $\bar w$:
$$
\lmat a & b \\ \bar b & \bar a \rmat  \lmat w \\ \bar w \rmat
=\lmat \sigma \\ \bar \sigma \rmat ,
$$
where
$$
a=1+\frac{\bar c}{\o} \text{sech} \,\alpha e^{-2x\sinh\,\alpha},
\quad
b=-\bar c \tanh\,\alpha e^{-2 x\sinh\,\alpha},
$$
and
$$
\sigma=-2i\o \bar c\bar\lambda\tanh\,\alpha e^{-2x\sinh\,\alpha}.
$$
The solution of this $2 \times 2$ system is
$$
w=\frac{\bar a \sigma-b  \bar \sigma}{\Delta},\tag6.10
$$
where $\Delta =|a|^2-|b|^2.$  To determine whether a global solution exists, we must
evaluate $\Delta$ with $c(t)$ given as above.

Taking $z=0$ in (5.8) we find
$$
\Im\l=\frac {\l -\bar \l}{2i}=-\frac{\Re w(x)}{\o}.\tag6.11
$$ 
Equations (6.10), (6.11) are coupled equations for $\l$ and $w$.

We note that the solitons for imaginary $p$ always have poles on the real axis. These occur at
the zeros of $\Delta$, hence, whenever $|a|=|b|$. To show that $\Delta$ always vanishes, first set $C=c_0e^{-2x\sinh\,\a}$,
so that $0<|C|<\infty$.
As $t$ varies, $|b|=|C\tanh\,\a|$ is constant, while $|a|$ ranges over the interval 
$(1-|C|e^{-\a}\sech\,\a,\,1+|C|e^{-\a}\sech\,\a)$.
Therefore $\Delta$ vanishes whenever (we may assume $\a>0$)
$$
1-|C|e^{-\a}\sech\,\a < |C|\tanh\,\a<1+|C|e^{-\a}\sech\,\a;
$$
this inequality simplifies to
$$
0<(e^{2\a}+1)(1-1/|C|)<4,
$$
which is always satisfied for a range of $|C|$ in $(0,\infty)$. Hence the soliton always
has poles for real $x$ and $t$.

\head 7. Conclusions \endhead

We have proved the existence of global solutions of the initial
value problems of the flows (1.2) and (1.3) under complementary conditions
on the scattering data. Two questions are often raised about these results.
Firstly, can the conditions on the initial data be expressed in terms of the 
initial values $u(x,0)$ and $w(x,0)$? Secondly, are the conditions sharp or
merely sufficient conditions for global existence?

As to the first question, it is typical in the theory of the scattering
transform 
that the picture is often simpler on the scattering side than on the 
potential, or spatial side. This is perfectly analogous to the features
of the Fourier transform. For example, under the Fourier transform,
convolution, a non-local operation, is mapped into pointwise multiplication,
a local operation. In the case of completely integrable systems,
action-angle variables are expressed very conveniently on the scattering side.
[NMPZ]. Thus, without saying that simple conditions for global existence
cannot be placed directly on $u$ and $w$, we should also not be surprised
that the solvability conditions are expressed so simply in terms of the
scattering data.

The second question is in fact one we wish to address in future work,
but we conjecture that the conditions are in fact sharp. One already sees 
in the example of the soliton in \S6 for the case of imaginary $p$: it has
poles for all time. We conjecture further that if one takes as initial data
reflection coefficients $r_\pm(\xi)$ with support on the wrong section
of $\Sigma$, then singularities will develop in finite time. For example,
we conjecture that if for real $p$ one chooses for $r_\pm(\xi)$ smooth functions with compact support
on the real line, singularities will develop in finite time. It would be interesting to use the
Riemann-Hilbert formulation to investigate the nature of the blow up of the
solution. Presumably it would develop a pole, since that is the nature of
the singularities of the solution of the inverse scattering problem.

The situation is in some respects similar to that in
the AKNS system, in which some kind of symmetry on the potential is needed
in order to guarantee global solvability. For example, the nonlinear Schr\"odinger equation is obtained for Hermitian or skew-Hermitian potentials, the
modified KdV equation is obtained for real symmetric or skew symmetric potentials. The symmetry of the potentials is related to the positivity of
the quadratic conservation laws. Without the symmetry, these conservation laws
are not positive definite; and global solutions of the unsymmetric AKNS systems
do not in general exist (cf. the discussion in [2].)

\head Appendix \endhead
 
\proclaim{Proposition A1} Let $p,\ q \in L^1(\Bbb R)$.
Then for $z\in \Cal U_1$ the wave function $\psi_+$ can be written in the form
$$
\psi_+(x,z)=\left(A_0(x,E)+\frac1z A_1(x,E)\right)e^{iEx},
\qquad
z\in \Cal U_1, \tag A1
$$
where $A_0,\  A_1$ are analytic in $\Im E>0$, 
continuous onto the real $E$-axis,
and
$$
A_0\sim \l^{-1},
\qquad
A_1\sim 0,
\qquad
E\api, \Im E>0.
$$
\endproclaim
The form of the wave functions in the other domains can be obtained
from the $s$ symmetry.
\demo{Proof} Using the identities
$$
z=2E-\frac1z,
\qquad
\frac{1}{z^2}=\frac{2E}{z}-1,
\qquad
k=E-\frac1z,
$$
and the {\it ansatz} (A1), the integral equation (2.3a) can be written
as the coupled system of Volterra integral equations
$$
\align
A_0(x,z)=&1 -\inxp \frac{1 - e^{- 2i E(x-y)}}{2i E}
\left((iEp(y)+q(y))A_0(y,z)+ipA_1(y,z)\right)\,dy,\\
A_1(x,z)=&-\inxp \frac{1 - e^{- 2i E(x-y)}}{2i E}
\left(( q(y)-2iEp(y))A_1(y,z)-ipA_0(y,z)\right)\,dy.
\endalign
$$
These may be solved by successive approximations for $\Im E>0$
in the standard way, and the analyticity of $A_0$ and $A_1$
established.
 Letting $E\api$ we obtain the limiting
equations for $A_0$ and $A_1$,
$$
A_0(x)=1-\frac12 \inxp p(y)A_0(y)\,dy,
\qquad
A_1(x)=\inxp p(y)A_1(y)\,dy,
$$
from which the result follows. \qed
\enddemo
\proclaim{Lemma A2} The wave function $\psi_+(x,z)$ has the representation
$$
\psi_+(x,z)=\l
\left(e^{iEx}+\inxp (K_0(x,y)+\frac1z K_1(x,y))e^{iEy})\,dy \right). 
\tag A2
$$
Similar representations hold for the other wave functions 
$\psi_-,\ \phi_{\pm}$. Here, $K_0$ and $K_1$ are smooth kernel functions, with properties
similar to the kernels in the Gel'fand-Levitan-Marchenko integral equations in the scattering
theory of the Schr\"odinger equation.
\endproclaim
{\sl Remark:} Kaup, ([10],  (3.7)) assumes the representation (A2) in his derivation of
the GLM equations.
\demo{Proof}These Fourier representations are proved by arguments similar
to those used in the standard GLM theory. Since $A_0$ and $A_1$ are analytic in the upper half $E$-plane,
and since $A_0-\l^{-1} \ap 0$ as $E\api$, they can be represented, 
for each $x$, as
Fourier transforms
$$
A_0=\l^{-1}+\int_{-\infty}^0H_0(x,s)e^{-iEs}ds,
\qquad
A_1=\int_{-\infty}^0 H_1(x,s)e^{-iEs}ds,
$$
Writing
$$
e^{iEx} A_0=  \l e^{iEx}+\int_{-\infty}^0 H_0(x,s)e^{iE(x-s)}ds,
$$
putting $y=x-s$, and changing variables in the integral, we obtain
$$
A_0e^{iEx}=\l\left( e^{iEx}+\inxp K_0(x,y)e^{iEy}dy\right),
$$
where $K_0(x,y)=\l^{-1} H_0(x,x-y)$. The other term is dealt with in the same way. \qed
\enddemo

\ms
{\bf Acknowledgements} The authors would like to thank Percy Deift
and Xin Zhou for helpful comments and suggestions. We also thank David Kaup
for useful conversations concerning some of the ideas in his paper. 

The first
author would also like to acknowledge the hospitality of the Mathematical
Sciences Research Institute, where many of the results in this paper were
obtained.

\head References \endhead

\flushpar{1.} Beals, R. and R. Coifman, "Scattering and inverse scattering for first
order systems," {\it Comm. Pure \& Applied Math.}, {\bf 37} (1984), 39-90.
\ss
\flushpar{2.} Beals, R. and R. Coifman, "Inverse scattering and
evolution equations," {\it Comm. Pure \& Applied Math.}, {\bf 38}
(1985), 29-42.
\ss
\flushpar{3.} Beals, R., P. Deift, C.Tomei, {\it Direct and Inverse Scattering on the Line},
Mathematical Surveys and Monographs, {\bf 28} American Math. Society,
Providence, Rhode Island, 1988.
\ss
\flushpar{4.} Camassa, R. and D. D. Holm, "An integrable shallow water equation with peaked solitons," {\it Phys. Rev. Letts.} {\bf 71} (1993), 1661-1664.
\ss
\flushpar{5.} Deift, P., S. Venakides, X. Zhou, ``The collisionless
shock
region for the long-time behavior of solutions of the KdV equation,''
{\it Comm. Pure Appl. Math.}, {\bf 47}, (1994), 199-206.
\ss
\flushpar{6.} Drazin, P.G. and R.S. Johnson, {\it Solitons: an introduction},
Cambridge University Press, Cambridge, 1989.
\ss
\flushpar{7.} Fordy, A.P. "Isospectral flows: their Hamiltonian
structures, Miura maps, and master symmetries," in {\it Solitons in
Physics, Mathematics, and Nonlinear Optics}, 97-121. P.J. Olver and D.H. Sattinger, eds. IMA Volumes in Mathematics and its Applications, {\bf 25}, Springer-Verlag, New York, 1990.
\ss
\flushpar{8.} Jaulent, M. ``On an inverse scattering problem with an energy dependent potential," {\it Ann. Inst. H. Poincar\'e, Sect. A}, (1972),
{\bf 17}, 363-372;  ``Inverse scattering problem in absorbing media,"
{\it J. Math. Phys.}, {\bf 17}, (1976), 1351-1360.
\ss
\flushpar{9.} Jaulent, M. and C. Jean, ``The inverse problem for the 
one-dimensional Schr\"od-
inger operator with an energy dependent potential, I, II."
{\it Ann. Inst. H. Poincar\'e, Sect. A}, {\bf 25}, (1976), 105-118;
119-137.
\ss
\flushpar{10.} Kaup, D. ``A higher-order water wave equation and the method of solving it," {\it Progr. Theoret. Phys.}, {\bf 54} (1975), 396-402.
\ss
\flushpar{11.} Matveev, V.B. et M.I. Yavor, "Solutions presque p\'eriodiques
et N-solitons de l'\'equation hydrodynamique non lin\'eaire de Kaup,"
{\it Ann. Inst. Henri Poincar\'e, Sect. A}, {\bf 31}, (1979), pp. 25-41.
\ss
\flushpar{12.} Novikov, S.P., S.V. Manakov, L.P. Pitatevksi,  V.E. Zakharov, {\it
Theory
of Solitons}, Consultants Bureau Publishing, New York, 1985.
\ss
\flushpar{13.} Sachs, R. ``On the integrable variant of the Boussinesq system:
Painlev\'e property, rational solutions, a related many-body system,
and equivalence with the AKNS hierarchy."
{\it Physica D} {\bf 30}, (1988), pp. 1-27;
 ``Polynomial $\tau$-functions for the
 AKNS hierarchy."
{\it Proceedings of Symposia in Pure Mathematics}, 
{\bf 49} (1989), Part 1, American
Math Society, 133-141.
\ss
\flushpar{14.} Sattinger, D.H. and J. Szmigielski, "Energy dependent scattering theory,"
{\it Differential and Integral Equations}, {\bf 8}, 1995, 945-959.
\ss
\flushpar{15.} Zakharov, V.E. and Shabat, A. B. ''Exact theory of two-dimensional self-
focusing and one-dimensional self-modulation of waves in nonlinear media,"
{\it Sov. Phys. JETP}, {\bf 34} (1972), 62-69.

\end